\documentstyle[aps,epsf,prl,multicol]{revtex}
\begin{document}
\draft
\title{Theoretical evidence for the semi-insulating character
of AlN}

\author{Antonella Fara,$^{(1)}$ Fabio Bernardini,$^{(1)}$ and
 Vincenzo Fiorentini$^{(1,2)}$} 
\address{(1)\, Istituto Nazionale per la Fisica della Materia and
Dipartimento di Fisica, 
Universit\`a di Cagliari, Italy\\
(2)\, Walter Schottky Institut, Technische Universit\"a{}t M\"u{}nchen, 
85748 Garching, Germany}
\date{\today}

\maketitle 
\begin{abstract}
We present ab initio density-functional  calculations 
for acceptors, donors, and  native
defects in aluminum nitride, showing that  acceptors are
deeper (Be $\sim$ 0.25 eV, Mg $\sim$ 0.45 eV)  and less soluble  than
in GaN; at further variance with GaN, both the extrinsic donors
Si$_{\rm Al}$ and C$_{\rm Al}$, and the native donor V$_{\rm N}$ (the
anion vacancy) are found to be  deep (about 1 to 3 eV below the
conduction). We thus predict that 
doped  AlN will generally turn out to be semi-insulating
in the normally achieved Al-rich
conditions, in  agreement with the known doping difficulties
 of high-$x$ Al$_x$Ga$_{1-x}$N alloys.
\end{abstract}

\begin{multicols}{2}
III-V nitrides have by now  established themselves as a key materials
system for high-frequency optoelectronics. Doping is still a central
problem in this area, especially as recent developments tend to focus
on heterojunction systems, and hence on alloys with large In and Al
content. Doping of either kind has been shown to be increasingly
difficult as the AlN fraction increases in Al$_x$Ga$_{1-x}$N alloys
 \cite{exp}. On the other hand, theoretical studies of doping  have
been much less detailed for AlN\cite{bogus,cathy,mattila}
than for GaN \cite{bogus,neug,noi,apl,neuprl}, and practically 
null for InN.   Here we
contribute to the discussion the results of  accurate
first principle density functional theory calculations for acceptors, 
donors, and vacancies in wurtzite AlN. We find that typical candidate
acceptors, namely cation-substituting  Be and Mg,  are moderately but
sizably deeper than in GaN. More interestingly, donors known to be
shallow in GaN (Si, C, and the nitrogen vacancy) are in fact deep in
AlN in their substitutional configuration.
Our results are in general  agreement with previous results,
when comparable. An exception is the  behavior of donors, which  is at
partial variance with previous studies \cite{bogus,cathy}, but seems
to agree with experimental data \cite{exp} indicating serious $n$-type
doping difficulties for AlN and high-$x$
Al$_x$Ga$_{1-x}$N alloys.  

\paragraph*{Method -- }
We use  local-density-functional-theory (LDA) \cite{dft}
 ultrasoft-pseudopotential plane-wave calculations  of  total energies
 and forces   in  doped AlN wurtzite supercells typically encompassing
32 atoms, and with plane wave cutoff of 25  Ry (for further technical
details see Refs. \cite{noi,apl,tesi}),  to predict from first
principles the  formation  energies and thermal ionization energies of
cation-substituting Be, Mg, Si, and C, and of the N and Al vacancies
in  AlN. The carriers concentration  at temperature T due to an impurity
or defect with thermal ionization energy $\epsilon$, and a  formation
energy   $E_{\rm form}$, is 
\begin{equation}  
n = N_s\ {\rm exp}\, (-E_{\rm form}/{\rm k_B  T_g}) \ 
{\rm  exp}\, (- \epsilon/{\rm k_B T}), 
\label{carrier}
\end{equation}
asssuming impurity incorporation or defect creation  in  thermal
equilibrium at a  growth temperature of T$_{\rm g}$, and with
$N_s$=2.44$\times$10$^{22}$ cm$^{-3}$  available cation or anion
sites (i.e. half the theoretical atomic density of AlN). Thus, the
largest the formation 
and
ionization energies, the less efficient the doping.
A non-zero formation entropy (neglected here) will of course enhance
the dopant concentration.  The formation energy for an
impurity in charge state {\it Q} is 
\begin{equation}
E_{\rm form} (Q) = 
E^{\rm tot}(Q)  - \sum_X n^{\rm X} \mu^{\rm X} + Q (\mu_e + E_v^Q),
\label{eform}
\end{equation}
with  $\mu_e$ the electron chemical potential
(synonimous with the Fermi energy E$_{\rm F}$ in our  
T=0 calculations),
$E^{\rm tot}(Q)$ the total energy of the fully-relaxed defected
supercell in  charge state $Q$, $E_v^Q$ its top valence band energy,
$n^{\rm X}$ and  $\mu^{\rm X}$
the number of atoms of the involved species (X=Al, N, impurity)
and their chemical potentials. The latter potentials  are determined
by the equilibrium conditions with AlN and the compounds (if any)
of the specific impurity  with Al or N. The structures and formation
enthalpies of the  solubility limiting compounds (Al$_2$O$_3$,
Si$_3$N$_4$, Be$_3$N$_2$, Mg$_3$N$_2$, and $d-$C) are calculated  
ab initio.  We generally assume the highest  $\mu^{\rm impurity}$  
compatible with the relevant  solubility limit. Then, only
 only one independent chemical potential is left (e.g. $\mu^{\rm N}$),
and its value is  determined by the imposed (N-rich $\leftrightarrow$
Al-rich) growth  conditions. 

The thermal ionization energy $\epsilon[0/-]$ of a single acceptor
is by definition the formation-energy difference of the two charge
states $Q$=0 and $Q$=--1 at $\mu_e$=0, and it corresponds to hole
release from (i.e., electron  promotion into) the  acceptor extrinsic
state with concurrent geometric relaxation.  For a single donor, the
analogous quantity is $\epsilon[+/0]$, the formation-energy difference
of the charge states $Q$=+1 and $Q$=0 at $\mu_e$=0. This quantity is
the distance of the energy level from the valence band. The thermal
ionization energy of the donor electron {\it into the conduction band}
is $E_{\rm gap} - \epsilon[+/0]$. Thus, the theoretical evaluation of
donor ionization energies with respect to the conduction edge requires
a decent estimate of the fundamental gap. The DFT-LDA 
eigenvalue gap is inappropriate for two reasons:
first, it is  notoriously inaccurate in general \cite{me}; second, it
seems inconsistent to compare the DFT-LDA eigenvalue-difference gap at
fixed number of electrons, with our ionization levels extracted from
total energy  differences of variable-electron-number systems. 

We thus proceed to evaluate the gap in analogy to impurity levels,
specifically as the ionization energy   $\epsilon[0/-]$ of a
defectless cell, i.e. the difference in total energy between bulk
supercells with N and N+1 electrons respectively. This way of
proceeding is technically consistent with  
the treatment of the defects, and the final result is affected by
essentially the same systematic errors: our result of 5.6 eV for
the gap,  within about 10 \%  of the experimental value of 6.2 eV,
indicates that the correct theoretical  gap to be compared to donor
levels is indeed somewhere near the experimental value. Yet another,  
conceptually more satisfactory estimate would be that dictated  by
$\Delta$SCF 
theory (see e.g. Ref. \cite{cappella}), whereby  the gap is the
difference  $\epsilon[0/-]$-- $\epsilon[+/0]$ for the defectless
cell. Using this expression,  we get 6.0 eV, even closer to
experiment. The  gap calculated as discussed above, and the ensuing
depth of the impurity levels, should be taken as   qualitative
indications \cite{nota}.  

\begin{figure}
\epsfclipon
\epsfysize=10cm
\centerline{\epsffile{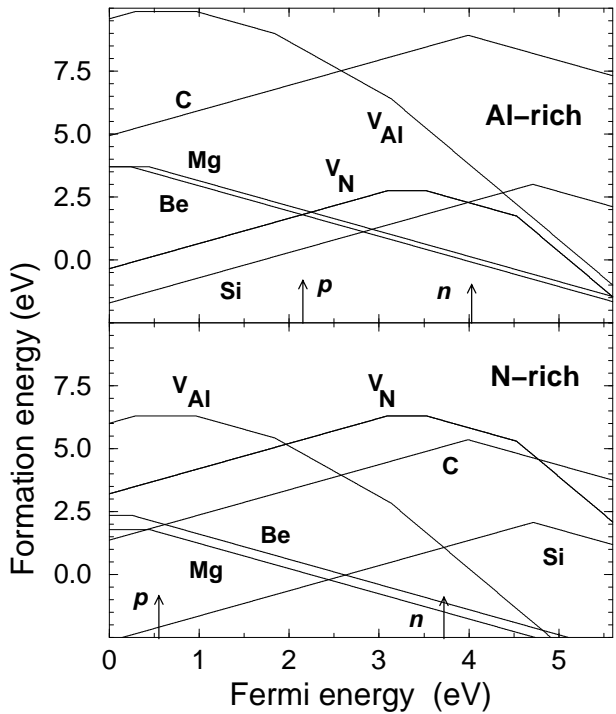}}
\narrowtext\caption{Formation energy of impurities and native defects
in AlN, for N-rich and Al-rich growth conditions. Arrows indicate the
best Fermi levels achievable in typical conditions (T$_g\sim$ 1200 
K).}
\label{fig.1}
\end{figure}

\paragraph*{Results -- } 
The results of our calculation are summarized in Figure \ref{fig.1},
for both Al-rich (upper panel) and N-rich (lower panel) growth
conditions. The thermal levels of  acceptors, referred to the valence
band top,  are 0.25 eV for Be$_{\rm
Al}$ and 0.45 eV for Mg$_{\rm Al}$. Thus, these candidate 
acceptors  are both appreciably deeper than in GaN, where the levels 
of Be$_{\rm Ga}$ and Mg$_{\rm Ga}$ are at about 0.1 and 0.2 eV,
respectively \cite{noi,apl}.

The extrinsic cation-substitutional donors are found to be
 deep, with  thermal  levels at 0.9 eV (Si) and 1.2 eV (C) below the
conduction band (the position of the latter is  the valence band top
position plus the gap estimated as discussed previously). 
Si and C are stable in the substitutional configuration,
and are found to be deep therein \cite{nota2}. 
Inspection of the partial charge density of the impurity
state of Si$_{\rm Al}$  confirms its strong localization. For both C 
and Si, the neutral state is never stable, i.e. both are negative-$U$
centers. The calculated $U$'s are -0.22 eV and -0.17 eV for Si and C
respectively. 

As for the vacancies, V$_{\rm Al}$ is a multiple acceptor  with a 
formation energy diving to zero and below in  $n$-type
conditions. This relatively unsurprising results parallels
closely that for the Ga vacancy in GaN \cite{neug}, and the
previous results on AlN by Mattila and Nieminen \cite{mattila}.
The nitrogen vacancy is a donor in $p$-type conditions, as in GaN:
however,  its first thermal level is very deep, indeed close to 
midgap.
Negative states of V$_{\rm N}$  are realized up to --3, in agreement
with previous results \cite{mattila}. We did not
investigate  the +2 and +3 states of V$_{\rm N}$ found in Ref. 
\cite{cathy}, but their  existance does not affect our 
conclusions. The realization of high charge states is not uncommon
 in insulators and wide gap semiconductors, a known example being
oxygen vacancies  in $\alpha-$quartz \cite{sio2}. 

The implications of these results for  doping are easily analyzed
using the theoretical framework by van de Walle {\it et al.}
\cite{vandewalle}. For our present purposes we just need to recall
that in the final analysys, in the
presence of competing donor and acceptor species, the Fermi level 
turns out to be pinned at about the intersection of the lowest
donor and acceptor formation energy curves. The Fermi levels 
obtained respectively via $n$ and $p$ doping  for typical realistic
conditions (T$_g\sim$ 1200 K, T$_{\rm oper} \sim$ 300 K)
are denoted by arrows in Figure \ref{fig.1}. Doping is clearly 
very inefficient in all cases; let us examine the reasons for that.

As for $p$-doping, it is  clear that in Al-rich conditions
the N vacancy compensates the acceptors, pinning the Fermi level
at about 2-2.5 eV above the valence band.   In the more favorable
N-rich conditions, this compensation is not effective (even if the
3+ state of Ref. \cite{cathy} is accounted for). However,
the formation and ionization energies of the acceptors are sizably
larger than in  GaN, and both the dopant and the carrier
concentration will be substantially lower than in GaN, making
$p$-doping of AlN even less efficient than in GaN. The shallowest
acceptor will produce in the most favorable case a Fermi level at
$\sim 0.5-0.6$ eV above the valence (Be$_{\rm Al}$  in N-rich
conditions at  a MOCVD-like  growth temperature T$_{\rm g}\simeq
1000-1200$ K). The co-incorporation mechanisms involving H
\cite{apl,neuprl} or  O \cite{apl,brandt} found to be effective in GaN
remain to be investigated in AlN.  

As for $n$-doping the situation is even worse. Firstly, the shallowest
level found here, that of Si$_{\rm Al}$ (the natural choice, as
Si$_{\rm Ga}$ is the standard shallow donor in GaN), is about 1 eV
below the conduction band (C has a huge formation energy and is always
unfavorable). Secondly,
vacancy compensation is always in action. The nitrogen vacancy will
compensate Si in Al-rich conditions, and the  Al vacancy will in
N-rich conditions.  Thus, in no event will the Fermi level be higher
than about 2 eV below the  conduction band, in  both
N-rich and Al-rich growth conditions. $n$-doping of AlN will
therefore be especially difficult, if at all possible.

In summary,  ab initio calculations predict that acceptors and donors
in use for GaN are not suitable for AlN and high-$x$ Al$_x$Ga$_{1-x}$N
 alloys.
 The acceptors Be and Mg
have high formation energies and fairly high ionization energies, and
suffer from nitrogen vacancy compensation in Al-rich conditions. The
donors Si and C are deeper than $\sim$ 1 eV below the conduction band,
and therefore unusable for doping of AlN. In addition, in
N- and  Al-rich growth conditions, respectively, the Al vacancy
and the N vacancy will act to compensate the donors. While the
acceptor behavior of the Al vacancy is unsurprising, the deep donor
character of the  N vacancy is at variance with the results for
GaN. In this respect our findings support the previous results by
Mattila and Nieminen   \cite{mattila} over those of Stampfl and van de
Walle \cite{cathy}.
While  moderate $p$-type doping might be obtained in
N-rich conditions, in Al-rich conditions AlN will be always
semiinsulating. 

V.F. acknowledges support by  the Alexander von Humboldt-Stiftung
during his stay at WSI.

\end{multicols}

\begin{references}
\bibitem{nota}
Since our calculated gap is   converged 
 with respect to cell size at the level of $\sim$ 5 \%,
it seems plausible to  draw the  semi-quantitative
conclusion that the total-energy difference gap is close to the
experimental gap.
\bibitem{nota2}
In our calculations we found no spontaneous DX distortion.
 Since we did not investigate possible
metastabilities (e.g. of donors) towards DX-like configurations,
we cannot exclude the latter can be realized (incidentally, 
none of the previous studies \protect\cite{bogus,cathy,mattila}
found any for Si or C). In any case, such metastability will not
improve  the shallowness of the levels and will therefore
 not change our conclusions.
\end{references}
\end{document}